\documentclass[9pt,sigconf,table]{acmart}

\usepackage[T1]{fontenc}
\usepackage[utf8]{inputenc}
\usepackage[nolist]{acronym}
\usepackage{url}
\def\UrlOrds{\do\*\do\-\do\~\do\'\do\"\do\-}%
\makeatletter
\g@addto@macro{\UrlBreaks}{\UrlOrds}
\makeatother
\usepackage{hyperref}
\usepackage{booktabs}
\usepackage{multirow}
\usepackage{color,soul}
\usepackage{balance}
\usepackage{tabularx}
\usepackage{xcolor}
\usepackage{blindtext}
\usepackage{mdframed}
\usepackage{subcaption}

\usepackage[misc]{ifsym}

\usepackage{graphicx}
\usepackage[export]{adjustbox}

\usepackage{tikz}
\newcommand\copyrighttext{%
	\footnotesize The final publication is available at the ACM DL via \url{https://dl.acm.org/doi/10.1145/3517745.3561445}}
	\newcommand\copyrightnotice{%
	\begin{tikzpicture}[remember picture,overlay]
	\node[anchor=south,yshift=10pt] at (current page.south) {\fbox{\parbox{\dimexpr\textwidth-\fboxsep-\fboxrule\relax}{\copyrighttext}}};
	\end{tikzpicture}%
}

\newcommand{\eg}{e.g., }

\newcommand{\ie}{i.e., }

\newcommand{\afblock}[1]{\noindent{\textbf{#1 }}}
\newcommand{\takeaway}[1]{\textit{\textbf{Takeaway:}} \textit{#1}}

\begin{acronym}
    \acro{EDNS}{Extended DNS}
    \acro{DoUDP}{DNS over UDP}
    \acro{DoTCP}{DNS over TCP}
    \acro{RA}{RIPE Atlas}
    \acro{DoT}{DNS over TLS}
    \acro{DoH}{DNS over HTTPS}
    \acro{DoQ}{DNS over QUIC}
    \acro{DoDTLS}{DNS over DTLS}
    \acro{GSO}{Generic Segmentation Offload}
    \acro{GRO}{Generic Receive Offload}
    \acro{DoS}{Denial of Service}
    \acro{QoE}{Quality of Experience}
    \acro{EU}{Europe}
    \acro{NA}{North America}
    \acro{AF}{Africa}
    \acro{SA}{South America}
    \acro{OC}{Oceania}
    \acro{AS}{Asia}
    \acro{PoP}{Point of Presence}
    \acro{UDP}{User Datagram Protocol}
    \acro{TCP}{Transmission Control Protocol}
    \acro{TLS}{Transport Layer Security}
    \acro{APNIC}{Asia Pacific Network Information Centre}
    \acro{RT}{Response Time}
    \acro{TFO}{TCP Fast Open}
    \acro{FQDN}{Fully Qualified Domain Name}
    \acro{RTT}{Round-Trip Time}
    \acro{ISP}{Internet Service Provider}
    \acro{RCODE}{Response Code}
    \acro{TTL}{Time to Live}
    \acro{CPE}{Customer-premises equipment}
    \acro{IAD}{Integrated access device}
    \acro{WG}{Working Group}
    \acro{IETF}{Internet Engineering Task Force}
    \acro{dnsop}{Domain Name System Operations}
    \acro{RR}{Resource Record}
    \acro{IP}{Internet Protocol}
    \acro{MTU}{Maximum Transmission Unit}
    \acro{RD}{Recursion Desired}
    \acro{ALPN}{Application-Layer Protocol Negotiation}
    \acro{IANA}{Internet Assigned Numbers Authority}
    \acro{SNI}{Server Name Indication}
    \acro{PLT}{Page Load Time}
    \acro{FCP}{First Contentful Paint}
\end{acronym}

\acrodefplural{AS}[ASes]{Autonomous Systems}
\acrodefplural{TLD}[TLDs]{Top-Level Domains}
\acrodefplural{TTFB}[TTFBs]{Times to First Byte}
\acrodefplural{PoP}[PoPs]{Points of Presence}
\acrodefplural{FQDN}[FQDNs]{Fully Qualified Domain Names}
\acrodefplural{RCODE}[RCODEs]{Response Codes}
\acrodefplural{RR}[RRs]{Resource Records}
\acrodefplural{RT}[RTs]{Response Times}



\copyrightyear{2022} 
\acmYear{2022} 
\setcopyright{acmcopyright}\acmConference[IMC '22]{Proceedings of the 22nd ACM Internet Measurement Conference}{October 25--27, 2022}{Nice, France}
\acmBooktitle{Proceedings of the 22nd ACM Internet Measurement Conference (IMC '22), October 25--27, 2022, Nice, France}
\acmPrice{15.00}
\acmDOI{10.1145/3517745.3561445}
\acmISBN{978-1-4503-9259-4/22/10}

\begin{document}
\title[Evaluating DNS over QUIC and its Impact on Web Performance]{DNS Privacy with Speed? Evaluating DNS over QUIC and its Impact on Web Performance}



\author{Mike Kosek}
\orcid{0000-0002-3299-5546}
\affiliation{%
  \institution{Technical University of Munich}
}
\email{kosek@in.tum.de}

\author{Luca Schumann}
\affiliation{%
  \institution{Technical University of Munich}
}
\email{schumann@in.tum.de}

\author{Robin Marx}
\affiliation{%
  \institution{KU Leuven}
}
\email{robin.marx@uhasselt.be}

\author{Trinh Viet Doan}
\orcid{0000-0003-4728-746X}
\affiliation{%
  \institution{Technical University of Munich}
}
\email{doan@in.tum.de}

\author{Vaibhav Bajpai}
\orcid{0000-0003-4089-1090}
\affiliation{%
  \institution{CISPA Helmholtz Center for Information Security}
}
\email{bajpai@cispa.de}


\renewcommand{\shortauthors}{M. Kosek et al.}

\begin{abstract}
    Over the last decade, Web traffic has significantly shifted towards HTTPS due to an increased awareness for privacy.
    However, DNS traffic is still largely unencrypted, which allows user profiles to be derived from plaintext DNS queries.
    While DNS over TLS (DoT) and DNS over HTTPS (DoH) address this problem by leveraging transport encryption for DNS, both protocols are constrained by the underlying transport (TCP) and encryption (TLS) protocols, requiring multiple round-trips to establish a secure connection.
    In contrast, QUIC combines the transport and cryptographic handshake into a single round-trip, which allows the recently standardized DNS over QUIC (DoQ) to provide DNS privacy with minimal latency.
    In the first study of its kind, we perform distributed DoQ measurements across multiple vantage points to evaluate the impact of DoQ on Web performance.
    We find that DoQ excels over DoH, leading to significant improvements with up to 10\% faster loads for simple webpages.
    With increasing complexity of webpages, DoQ even catches up to DNS over UDP (DoUDP) as the cost of encryption amortizes: With DoQ being only $\sim$2\% slower than DoUDP, encrypted DNS becomes much more appealing for the Web.
    \copyrightnotice

\end{abstract}

\maketitle


\section{Introduction}
\label{sec:introduction}

Led by the increased awareness for Internet security and privacy, HTTPS has replaced HTTP and became the default Web protocol over the last decade~\cite{tls.adoption,httpsusage}.
However, despite the DNS being one of the most crucial parts of the Internet infrastructure, unencrypted DNS traffic using \ac{DoUDP} and \ac{DoTCP} is still the norm~\cite{deccio2019dns}.
Hence, even with the encryption of the actual Web content, browsing behaviors and user profiles can still be derived and even tracked by observing unencrypted DNS queries~\cite{KimZ15,KirchlerHLK16,LiMGLZLG18,connection.oriented.dns}.
This problem was originally addressed by the encrypted protocols \ac{DoT}~\cite{rfc7858} and \ac{DoH}~\cite{rfc8484}, which have been integrated by browsers and public DNS resolvers since 2016~\cite{dns.privacy.resolvers, cf.dot.doh, google.dot, google.doh}.
As these protocols have been extensively studied in terms of response times~\cite{doan2021measuring,encrypted.dns.be.fast,cuadrado2019dns,lu2019dns,lu2019dns,doh.round.the.world} and impact on Web performance~\cite{hounsel2020dns,cuadrado2019dns,borgolte2019dns}, it has become clear that both DoT and DoH are constrained by the round-trips required for the handshakes of the underlying transport (TCP) and encryption (TLS) protocols.


To overcome these limitations, the QUIC transport protocol~\cite{rfc9000}, standardized in 2021, combines the transport and cryptographic handshake into a single round-trip.
Consequently, \ac{DoQ}~\cite{rfc9250} aims to provide DNS privacy with minimal latency.
While it was only recently standardized in May 2022, DoQ is already deployed by privacy-focused DNS resolvers in production systems~\cite{adguard,nextdns}.
However, as of September 2022, only one study focusing on DoQ exists:
In our preliminary work, we compared DNS protocol performance measured from a single vantage point~\cite{2022_pam_doq}.
We showed that the adoption of DoQ by public DNS resolvers is slowly increasing and that although DoQ outperforms DoT and DoH in terms of DNS \emph{single query response time}, around 40\% of measurements still result in considerably slower response times than expected due to the enforcement of QUIC's traffic amplification limit~\cite{rfc9000}.


To advance this state of the art, we (1) perform distributed measurements across 6 vantage points and (2) add support for TLS 1.3 \texttt{Session Resumption} and \texttt{0-RTT} for DoQ, DoT, and DoH.
While we find that no public resolver supports \texttt{0-RTT}, our measurements are not constrained by QUIC's traffic amplification limit due to \texttt{Session Resumption}, which can therefore significantly improve the \emph{single query response time}: DoQ outperforms DoT and DoH by $\sim$33\%, making encrypted DNS much faster.
We further conduct \emph{Web performance} measurements to analyze and compare the impact of DoQ on Web browsing.
In our distributed measurements, we find that DoQ significantly improves over DoH by up to 10\% faster loads for simple webpages.
With increasing complexity of webpages, DoQ even catches up to DoUDP as the cost of encryption amortizes the more DNS queries are required for loading a webpage: with DoQ being only $\sim$2\% slower than DoUDP, encrypted DNS becomes much more appealing for the Web.
\section{Methodology}
\label{sec:methodology}

To study the response times of \ac{DoQ} in comparison to \ac{DoUDP}, \ac{DoTCP}, \ac{DoT}, and \ac{DoH} for single queries and assess their impact on Web performance, we issue distributed measurements using 6 vantage points while targeting 313 DNS resolvers worldwide.

\textbf{Target Resolvers and Vantage Points.}
To identify DoQ resolvers, we issue a scan of the IPv4 address space in week 14 of 2022 from a single vantage point located in the research network of the Technical University of Munich, Germany, targeting all proposed DoQ ports (\texttt{UDP} \texttt{784}, \texttt{853}, and \texttt{8853}~\cite{ietf-dprive-dnsoquic,rfc9250}).
For this, we leverage the \emph{ZMap}~\cite{the_zmap_project} network scanner, probing with a QUIC \texttt{INITIAL} packet with an invalid version number of 0:
By receiving a \texttt{Version Negotiation} packet in response, we identify the IP addresses that support QUIC on the respective port, without creating state on the target~\cite{rfc9000} to avoid exhausting resources. 
We then establish a connection to the identified targets, offering the DoQ \ac{ALPN} identifiers~\cite{ietf-dprive-dnsoquic,rfc9250}; if the connection establishment is successful, the target is identified to support DoQ~\cite{doq_verify}.

\begin{figure}[t]
	\centering
	\includegraphics[width=\linewidth,trim=2cm 1.54cm 1cm 0.4cm, clip, frame]{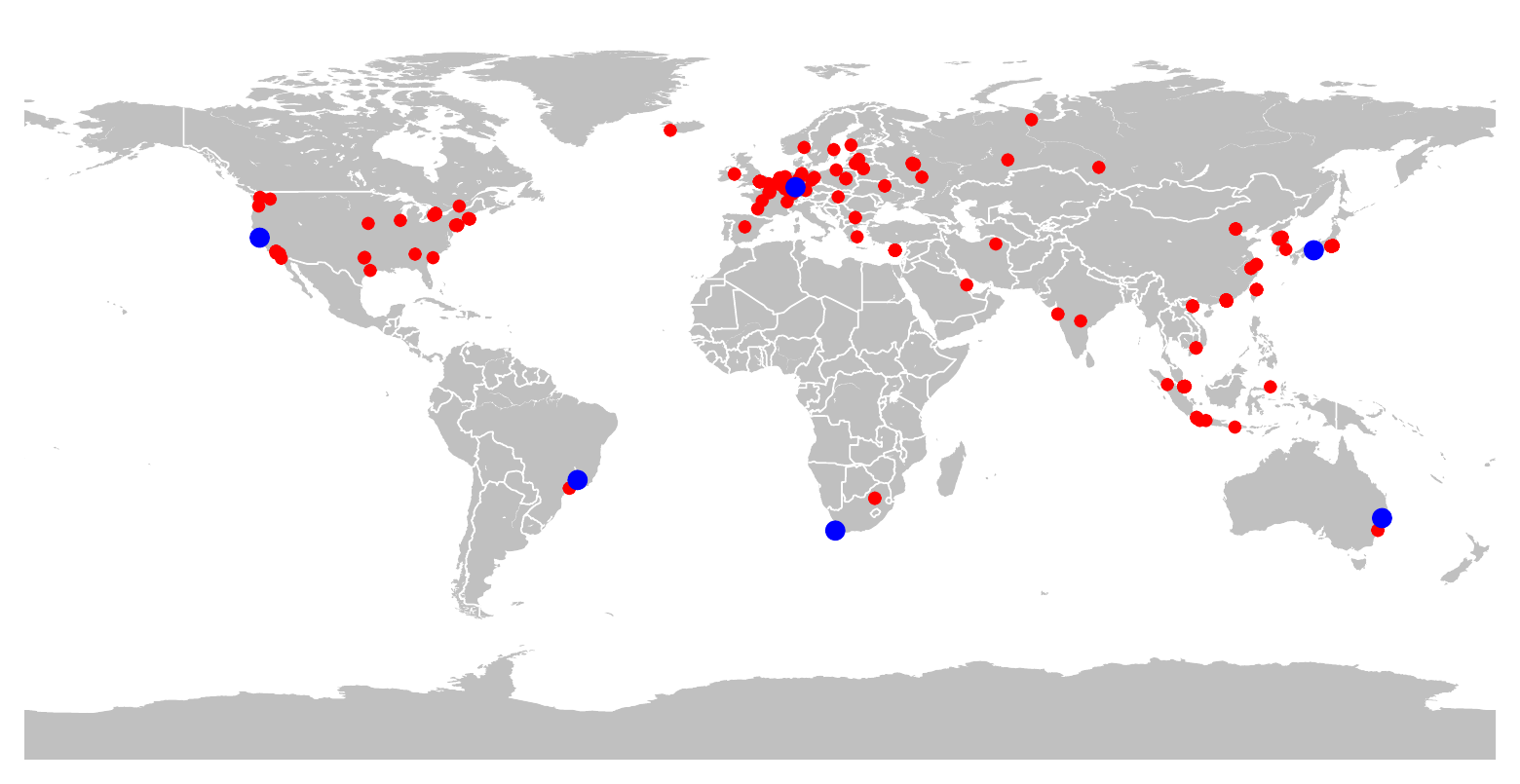}
	\vspace{-2em}
	\captionof{figure}{Geographical distribution of the 313 verified DoX resolvers (red dots) and vantage points (blue dots).
	} 
	\label{fig:resolvers-vps-map}
\end{figure}

Using this methodology, we identify 1,216 DoQ resolvers.
Comparing this finding with our preliminary work, which identified 1,217 DoQ resolvers in week 03 of 2022~\cite{2022_pam_doq}, we observe that the adoption of DoQ is currently stagnating. 
To enable a comparison of DoQ to the established DNS protocols, we further check the identified DoQ resolvers for their support of \ac{DoUDP}, \ac{DoTCP}, \ac{DoT}, and \ac{DoH}.
For this, we optimistically query the resolvers using \emph{DNSPerf}, an open-source DNS measurement tool supporting all stated protocols~\cite{doq_dnsperf}.
Of the 1,216 identified DoQ resolvers, we find that 548 support DoUDP, 706 DoTCP, 1,149 DoT, and 732 DoH, while their full intersection (i.e, resolvers supporting every DNS protocol) results in 313 verified DoX resolvers (preliminary work: 264~\cite{2022_pam_doq}).
While we acknowledge that public DNS resolvers often leverage IP anycast, we cross-reference anycast IP addresses used in related work~\cite{doan2021measuring,lu2019dns,doh.round.the.world,hounsel2020dns,borgolte2019dns,dotcp}, although without finding an overlap.
\autoref{fig:resolvers-vps-map} (red dots) presents the geographical distribution of the verified DoX resolvers based on an IPv4 geolocation lookup service~\cite{ipapi}, for which we observe that the majority are located in \ac{EU} with 130 resolvers, followed by \ac{AS} with 128, \ac{NA} with 49, and \ac{AF}, \ac{OC}, and \ac{SA} with 2 resolvers each.
Moreover, we find that the resolvers are distributed over 107 Autonomous Systems, with the majority located in ORACLE (47, 15.0\%), DIGITALOCEAN (20, 6.4\%), MNGTNET (18, 5.8\%), and OVHCLOUD (16, 5.1\%). 
The remaining Autonomous Systems each host 12 or less resolvers.
All measurements are performed using 6 distributed \emph{Amazon EC2} instances (\autoref{fig:resolvers-vps-map}, blue dots), employing one vantage point per continent.

\textbf{Single Query Response Time and Size.}
To study the single query response times and sizes of DoQ in comparison to \ac{DoUDP}, \ac{DoTCP}, \ac{DoT}, and \ac{DoH}, we leverage the open-source DNS measurement tool \emph{DNSPerf}~\cite{doq_dnsperf}.
Targeting the 313 verified DoX resolvers, we issue \emph{single query} measurements for all stated protocols on every vantage point, repeated every 2 hours, over the course of week 16 of 2022.
For this, an \texttt{A} record for \texttt{google.com} is queried.
We precede every measurement with an identical cache warming query to ensure that the following \emph{actual} measurement is directly answered from a cached record at the resolver, which avoids inconsistencies in the measured response times caused by recursive lookups.
This further allows us to reuse the TLS session parameters of the cache warming for the \emph{actual} measurement of DoQ, DoT, and DoH: By adding support for TLS 1.3 \texttt{Session Resumption} and \texttt{0-RTT} to \emph{DNSPerf}, we advance the state of the art of our preliminary work~\cite{2022_pam_doq} (which does not consider either feature) and by default use both mechanisms if supported by the resolver.
Additionally, we also store the negotiated \texttt{QUIC Version} as well as the \texttt{Address Validation} token received in a \texttt{New\_Token} frame of the cache warming query.
Reusing these in the QUIC \texttt{INITIAL} packet of the \emph{actual} DoQ measurement ensures that the QUIC handshake is not influenced by its \texttt{Version Negotiation} or \texttt{Address Validation} mechanisms, which would otherwise add 1 RTT each~\cite{2022_pam_doq}.
Hence, our DoQ implementation follows the recommendations of the DoQ standard, stating that \texttt{Address Validation} tokens should only be used in union with \texttt{Session Resumption}~\cite{rfc9250}.
Altogether, our methodology enables comparable response time measurements of a typical DNS usage scenario for all protocols for the first time, where a session between a client and a resolver is established to perform a single DNS query.

\textbf{Web Performance.}
To assess the impact of DoQ on Web performance in comparison to \ac{DoUDP}, \ac{DoTCP}, \ac{DoT}, and \ac{DoH}, we develop an open-source framework using \emph{Selenium}~\cite{selenium}, \emph{Chromium} \cite{chromium}, as well as \emph{DNS Proxy}~\cite{dnsproxyjustus}.
Using this framework, we issue Web performance measurements targeting the top 10 most popular webpages from the research-oriented \emph{Tranco} top list~\cite{tranco} as of April 12, 2022.
For this, we load every webpage using each DNS protocol via every one of the 313 verified DoX upstream resolvers from all vantage points, repeated every 48 hours, over the course of week 16 of 2022.
For each measurement, \emph{DNS Proxy} is newly setup as Chromium's local resolver on the \emph{Amazon EC2} instances and configured to forward the queries to the upstream DoX resolver by using either DoQ, \ac{DoUDP}, \ac{DoTCP}, \ac{DoT}, or \ac{DoH}.
The local DNS caches of both the operating system and \emph{DNS Proxy} are disabled to ensure that queries are forwarded to the configured upstream resolver.
In the next step, we leverage \emph{Selenium} to launch \emph{Chromium} and navigate to each webpage twice in succession:
As with the DNS \emph{single query}, the first navigation populates the upstream resolver's cache and ensures that the DNS queries of the second, \emph{actual} measurement navigation are directly answered from that resolver's cache.
As with \emph{DNSPerf}, we extend \emph{DNS Proxy} to support TLS 1.3 \texttt{Session Resumption} and \texttt{0-RTT}, and track the negotiated QUIC versions and tokens.
As such, this approach is identical to the aforementioned \emph{single query} measurements, following the recommendations of the DoQ standard~\cite{rfc9250}.
After the cache warming navigation, all sessions of \emph{DNS Proxy} are reset to ensure that a new session to the resolver is established for the \emph{actual} Web performance measurement, where TLS \texttt{Session Resumption} and \texttt{0-RTT} are used if supported by the DoX resolver.
Overall, this methodology allows us to compare DoQ with \ac{DoUDP}, \ac{DoTCP}, \ac{DoT}, and \ac{DoH} regarding their impact on Web performance for the first time, representing a typical usage scenario in which multiple DNS queries are sent when visiting a webpage.

\textbf{Ethical Considerations.}
To adhere to ethical principles and minimize the impact of our measurements, we respectfully follow best practices of the Internet measurement community~\cite{zmap.ethics,ethics.in.network.mm}:
We restrict our Internet-wide scans for the verification of DoX resolvers to one week and one vantage point in order to limit the traffic sent to the network operators' infrastructure. 
To allow targets to opt-out from our measurements, we display contact information and a description about the intent of our measurements on a webpage reachable via the IP address of each vantage point.
Further, we only target publicly reachable IP addresses and exclude targets based on a blocklist which is maintained across research groups within our University, and, therefore, also covers targets excluded from previous measurement studies.

\textbf{Reproducibility and Community Contributions.}
In order to enable the reproduction of our findings~\cite{reproducibility}, we make the developed tools, the raw data of our measurements, and the analysis scripts publicly available~\cite{doq_web_perf_open_source}.
Moreover, we upstream our changes to the tools used in our measurements as outlined in this chapter, aiming to facilitate future DNS protocol studies~\cite{doq_dnsperf_pr,doq_dnsmeasurements_pr,doq_dnsproxy_pr_1,doq_dnsproxy_pr_2}.
\section{Evaluation}
\label{sec:evaluation}

We begin the evaluation with an overview of the measurements, followed by the analysis of the \emph{single query} measurements in \autoref{sec:response-times}.
In \autoref{sec:web-performance}, we detail our findings on the impact of DoQ on \emph{Web performance} in comparison to \ac{DoUDP}, \ac{DoTCP}, \ac{DoT}, and \ac{DoH}.




Analyzing DoQ, we find that 89.1\% of all measurements are performed using QUIC version \texttt{1}.
The remaining measurements use the older QUIC \texttt{draft} versions \texttt{-34} (8.5\%), \texttt{-32} (1.8\%), and \texttt{-29} (0.6\%).
We find no differences between the QUIC versions, which confirms our expectations, as all observed versions are feature equivalent.
While our tooling supports all available DoQ versions as of April 18, 2022 (\emph{doq} for the standard~\cite{rfc9250}, as well as \emph{doq-i00} to \emph{doq-i11} for the \texttt{draft} versions), we find that \emph{doq-i02} is used in the majority of measurements with 87.4\%, followed by \emph{doq-i03} with 10.8\%, and \emph{doq-i00} with 1.8\%.
While the observed versions \emph{doq-i00} and \emph{doq-i02} are feature equivalent, \emph{doq-i03} changed the QUIC stream mapping to include a 2 byte message length field in order to permit multiple responses for a single query.
As with QUIC, however, we find no differences between DoQ versions.
All DoQ measurements use TLS 1.3 as mandated by the QUIC standard~\cite{rfc9001}.
As for DoT and DoH, around 99\% of measurements are performed using TLS 1.3, whereas the remainder use TLS 1.2.
Moreover, all DoH measurements use HTTP/2.
For both the QUIC-based DoQ and the TCP-based DoH and DoT, we find that no resolver supports TLS 1.3 \texttt{0-RTT}.
However, all resolvers support TLS 1.3 \texttt{Session Resumption} and respond with the maximum session ticket lifetime of 7 days as defined by the standard~\cite{rfc8446}; hence, \texttt{Session Resumption} is used in all TLS 1.3 measurements.
Lastly, we find that no resolver supports \texttt{\acl{TFO}} (TFO)~\cite{rfc7413} or \texttt{edns-tcp-keepalive}~\cite{rfc7828}.

\subsection{Single Query Response Time and Size}
\label{sec:response-times}

\begin{table}[t]
    \caption{Median single query sizes in bytes, as well as sample sizes of single query response time and Web performance measurements.}
    \vspace{-1em}
    \label{tab:dataset-overview}
    \resizebox{\linewidth}{!}{%
    \begin{tabular}{lrrrrr}
        \toprule
        {} & DoUDP & DoTCP & DoQ & DoH & DoT \\
        \midrule
        \multicolumn{4}{l}{\textbf{\begin{tabular}[c]{@{}c@{}}Single Query Sizes (median IP payload in bytes)\end{tabular}}} \\
        \rowcolor[gray]{.9}
        --- \texttt{Total} & 122 & 382 & 4444 & 2163 & 1522 \\
        --- \texttt{Handshake C-->R} & --- & 72 & 2564 & 569 & 551 \\
        \rowcolor[gray]{.9}
        --- \texttt{Handshake R-->C} & --- & 40 & 1304 & 211 & 211 \\
        --- \texttt{DNS Query} & 59    & 149 & 190 & 579 & 261 \\
        \rowcolor[gray]{.9}
        --- \texttt{DNS Response} & 63 & 121 & 386 & 804 & 499 \\
        \midrule
        \multicolumn{4}{l}{\textbf{\begin{tabular}[c]{@{}c@{}}Single Query Response Time\end{tabular}}} \\
        \rowcolor[gray]{.9}
        --- \texttt{Samples} & 154,092 & 154,503 & 159,676 & 157,637 & 158,959 \\
        \midrule
        \multicolumn{4}{l}{\textbf{\begin{tabular}[c]{@{}c@{}}Web Performance\end{tabular}}} \\
        \rowcolor[gray]{.9}
        --- \texttt{Samples} & 57,032 & 56,428 & 57,393 & 56,840 & 56,440 \\
        \bottomrule
    \end{tabular}
    }
\end{table}

The \emph{single query} measurements reflect a typical DNS usage scenario, where a session between a client and a resolver is established in order to perform a single DNS query.
\autoref{tab:dataset-overview} presents the sample sizes for the \emph{single query} measurements, where we observe slight variations due to resolvers not responding to every DNS query.
In our analysis, we differentiate between the \emph{handshake time}, the \emph{resolve time}, and the \emph{sizes} to account for the different transport protocol mechanisms used by the measured protocols.

\afblock{Handshake time.}
We define the \emph{handshake time} as the time between the client sending the first packet of the transport protocol handshake until the (encrypted) session to the resolver is established.
Thus, DoUDP is excluded from this analysis since UDP is connectionless.
As no resolver supports \acs{TFO} (see~\autoref{sec:evaluation}), the DoTCP handshake is expected to complete within 1 round-trip due to the TCP 3-way handshake.
For the encrypted protocols DoQ, DoH, and DoT, we find that no resolver supports \texttt{0-RTT}, but all resolvers support \texttt{Session Resumption} (see~\autoref{sec:evaluation}).
Hence, the DoQ handshake is also expected to complete within 1 round-trip due to QUIC's combination of the transport and encryption handshake.
Because DoH and DoT leverage TCP and no resolver supports \acs{TFO}, the transport and encryption handshake is expected to take 2 round-trips when TLS~1.3 is used (around 99\% of measurements, see ~\autoref{sec:evaluation}), and 3 round-trips in case of TLS~1.2.

\begin{figure}[!h]
    \centering
	\includegraphics[width=\linewidth]{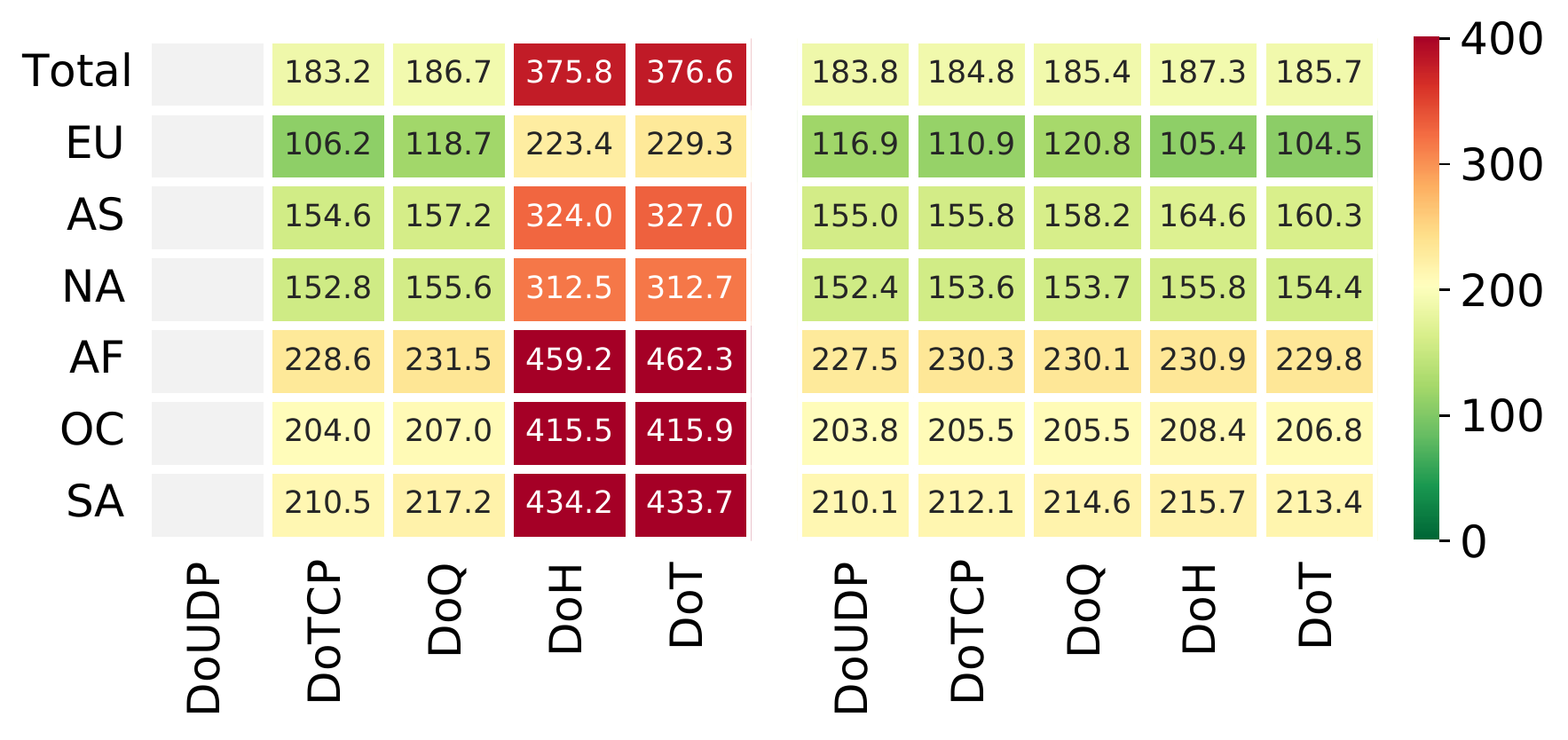}
    \raisebox{5ex}{
        \subfloat[\label{fig:handshake-resolve-heatmap-a} \emph{Handshake time}.]{\hspace{.5\linewidth}}
        \subfloat[\label{fig:handshake-resolve-heatmap-b} \emph{Resolve time.\phantom{-----------------------------}}]{\hspace{.6\linewidth}}
    }
    \vspace{-2em}
    \caption{Median \emph{Handshake time} (a, left) and \emph{Resolve time} (b, right) in ms per protocol over all vantage points (top row) and per vantage point (bottom rows). Ordered by the number of verified DoX resolvers per continent.}
	\label{fig:handshake-resolve-heatmap}
    \vspace{-0.5em}
\end{figure}




The median \emph{handshake times} in ms per protocol and vantage point are presented in \autoref{fig:handshake-resolve-heatmap-a}.
We find that both DoH and DoT show comparable \emph{handshake times} as expected, with the median over all vantage points (\emph{Total}, top row) being $\sim$376ms for DoH and $\sim$377ms for DoT. 
In comparison, both DoTCP and DoQ result in roughly half of that with $\sim$183ms for DoTCP and $\sim$187ms for DoQ, again confirming our expectations.
While our preliminary work~\cite{2022_pam_doq} found the \emph{handshake times} of DoQ to be considerably slower in comparison to DoTCP, this was due to DoQ being limited by QUIC's traffic amplification limit: The handshake is prolonged by 1 round-trip if the X.509 certificate offered by the resolver does not fit into the traffic amplification limit of three times the amount of data the resolver received in the \texttt{INITIAL} from the client (at least 1,200 bytes, see \emph{Sizes} in~\autoref{sec:response-times}).
In contrast, the measurements presented are not constrained by this limit due to the usage of \texttt{Session Resumption} (see~\autoref{sec:methodology}), where the X.509 server certificate is not exchanged yet again.


\afblock{Resolve time.}
We define the \emph{resolve time} as the time between the client sending the first packet of the DNS query until a valid DNS response is received.
As we ensure that the queried DNS record is cached on the target resolver (see~\autoref{sec:methodology}), the \emph{resolve times} are expected to be similar for all measured protocols if the protocols are handled equally by the path.

The median \emph{resolve times} in ms per protocol and vantage point are presented in \autoref{fig:handshake-resolve-heatmap-b}, where we find that all protocols indeed result in fairly similar \emph{resolve times} on each vantage point (\ie row-wise), and also over all vantage points (\emph{Total}, top row).
Moreover, we observe that the \emph{resolve times} correlate with the locations of the vantage points and resolvers (see~\autoref{fig:resolvers-vps-map}): While \ac{AF}, \ac{OC}, and \ac{SA} host only 2 resolvers each (bottom rows), we observe the highest median \emph{resolve times} on these vantage points due to larger geographical distances to the targeted resolvers.
In turn, \ac{AS} (128 resolvers) and \ac{NA} (49 resolvers) show significantly faster median \emph{resolve times}, only being outperformed by \ac{EU} where we find the highest geographical density of DoQ resolvers (130 resolvers).

\afblock{Sizes.}
There is a secondary general performance axis besides response times, namely size overhead.
While DoUDP adds just a few 8-byte UDP packet headers, DoQ incurs a heavy cost due to its modern handshake.
Table~\ref{tab:dataset-overview} shows the median incoming/outgoing IP payload bytes per protocol for a single \texttt{A} query for \texttt{google.com} to all DoX resolvers (using \texttt{Session Resumption} where possible), split over the handshake (Client->Resolver and vice versa) and actual DNS query/response.

We find that the encrypted protocols indeed transfer significantly more bytes than DoUDP and DoTCP.
The bytes transferred for a single DoQ handshake then again more than doubles in comparison to DoH and DoT, as QUIC requires all its \texttt{INITIAL}-carrying datagrams to be padded to at least 1,200 bytes to ensure MTU allowance on the network~\cite{rfc9000}.
However, DoQ's query and response sizes are relatively small in comparison to DoH due to the HTTP/2 overhead of DoH, e.g., message framing and header compression setup.
Hence, our results indicate that re-using a QUIC connection for multiple queries will mitigate its up-front cost faster than DoH. 




\takeaway{
Our findings on single query response time and size emphasize the importance of standards and their implementations: \texttt{Session Resumption} can significantly catalyze DoQ, outperforming DoT and DoH by $\sim$33\%.
Hence, DoQ makes encrypted DNS much more appealing than DoH, where DoQ falls short of DoUDP by only $\sim$50\% (DoT and DoH: $\sim$66\%) for single queries.
Consequently, DoQ's roughly double handshake size overhead over DoH seems a small price to pay.
}

\subsection{Web Performance}
\label{sec:web-performance}

Applying our methodology of the \emph{single query} measurements to \emph{DNS Proxy} (see~\autoref{sec:methodology}), we issue \emph{Web performance} measurements targeting the top 10 most popular webpages from the \emph{Tranco} list~\cite{tranco}.
To avoid initial redirects to the actual landing page (i.e., \emph{\nolinkurl{google.com}} will redirect to \emph{\nolinkurl{www.google.com}}), we replace the URLs by the actual landing page to ensure comparable results.
Representing a typical usage scenario where multiple DNS queries are sent when visiting a webpage, we analyze two standard Web performance metrics: \emph{\acl{FCP}} (FCP) and \emph{\acl{PLT}} (PLT).
\acs{FCP} marks the moment when the very first visible image or text is shown on the screen~\cite{mozilla-fcp}.
For \acs{PLT}, we calculate the time difference between the very start of the page load (corresponding to the start of the first DNS query/connection) and the \texttt{onLoad} event (corresponding to the moment when the webpage has finished loading), i.e., \texttt{LoadEventStart}$-$\texttt{NavigationStart}~\cite{mozilla-plt}.
FCP occurs early in the page load and, thus, should be more impacted by the DNS response times than PLT, which occurs late and is influenced by many other Web performance aspects.
\autoref{tab:dataset-overview} presents the sample sizes for the \emph{Web performance} measurements, where we again observe slight variations due to resolvers not responding to every DNS query.

\begin{figure}[t]
    \centering
	\subcaptionbox{\emph{First Contentful Paint}\label{fig:udp-baseline-fcp}}
		 {\includegraphics[width=.49\linewidth]{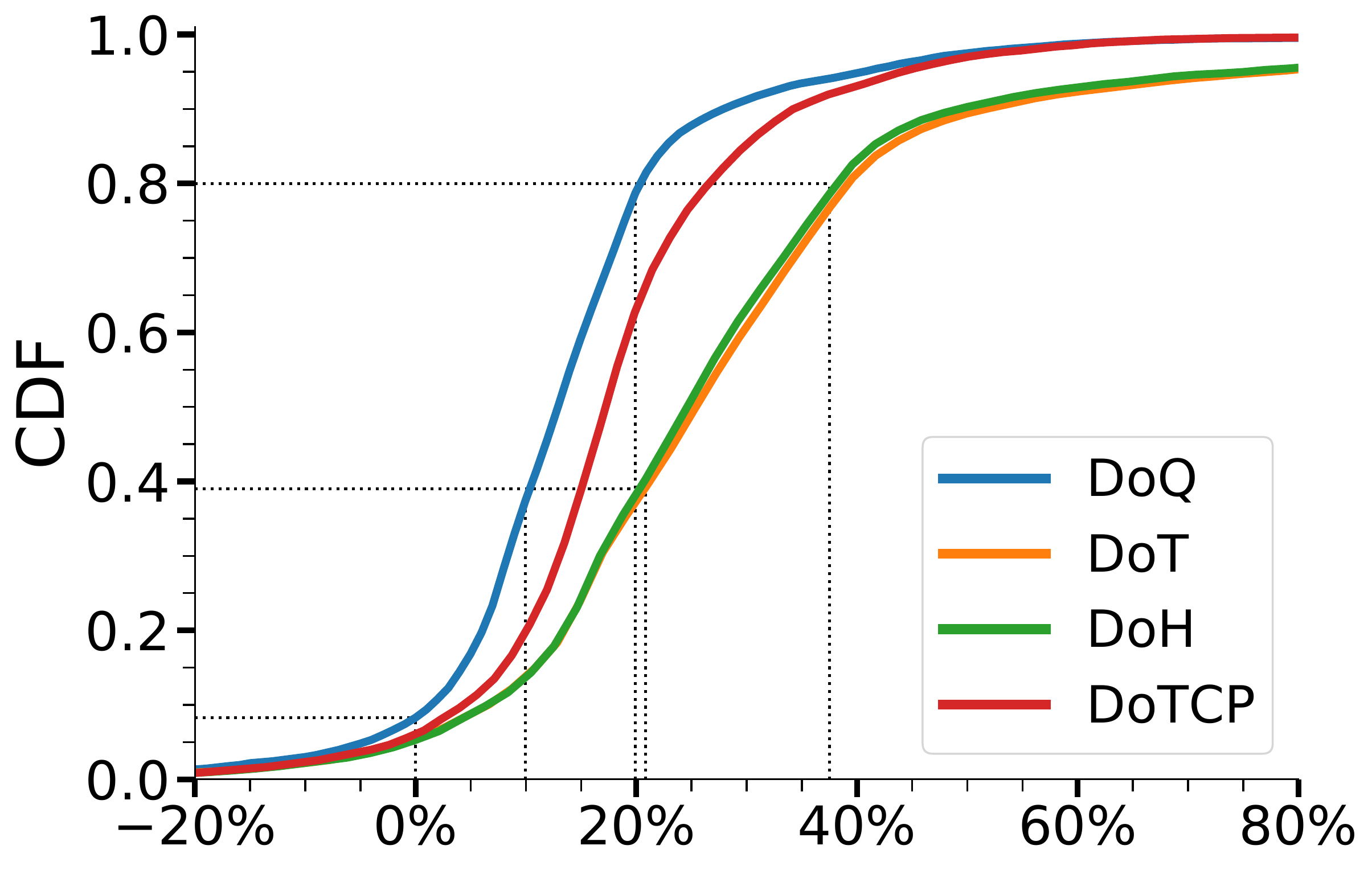}}
	\subcaptionbox{\emph{Page Load Time}\label{fig:udp-baseline-plt}}
		{\includegraphics[width=.49\linewidth]{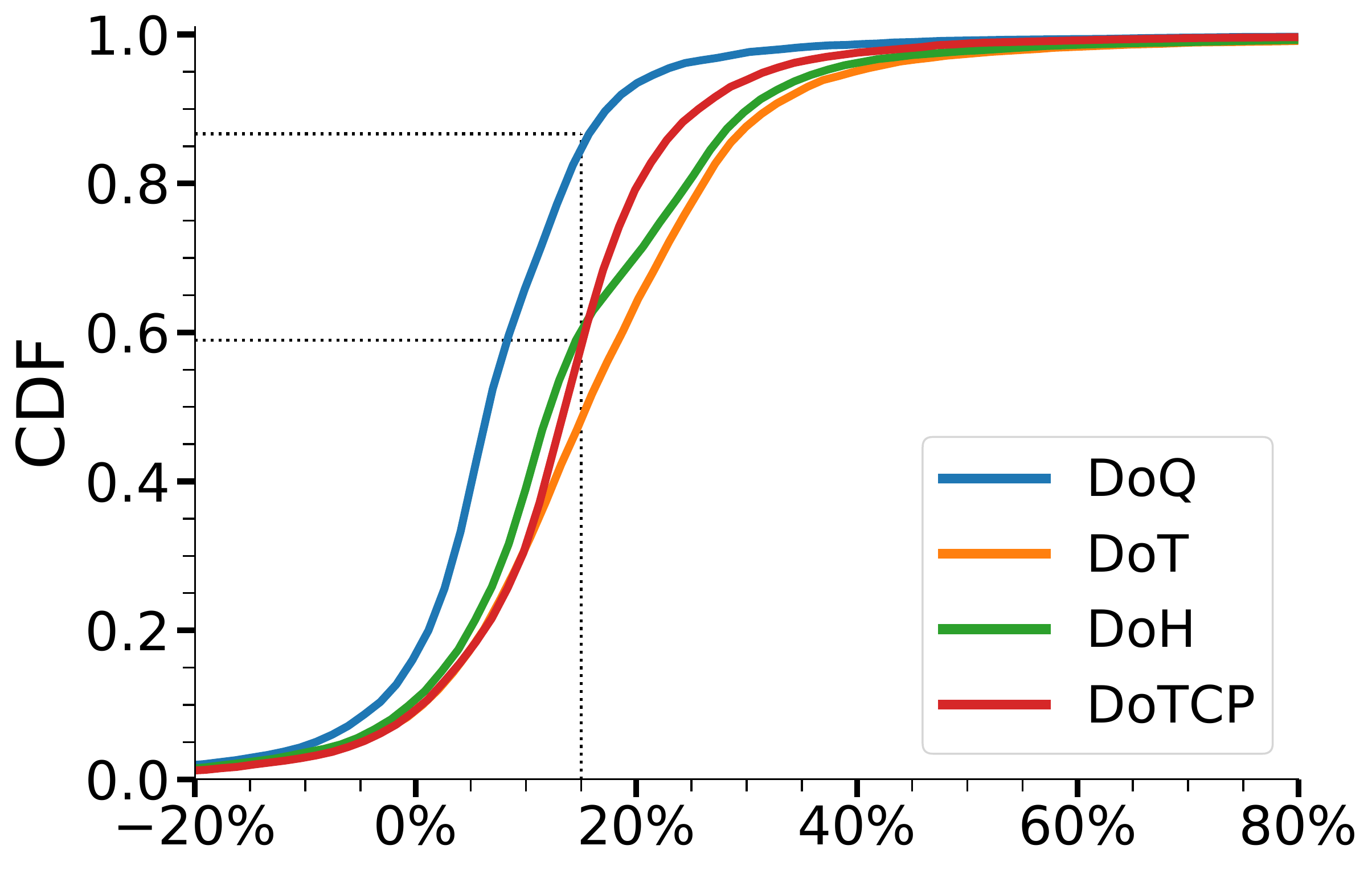}}
	\vspace{-0.5em}
	\caption{CDFs of the relative differences in FCP and PLT between DNS protocols with DoUDP as baseline.}
	\label{fig:udp-baseline}
\end{figure}

For each \texttt{[vantage point:resolver:DNS protocol]} combination, we perform four measurements using cold start page loads for every webpage.
We determine the medians of these measurements, enabling the comparison of upstream resolvers with different response times (i.e., to account for different geographical distances of vantage point and resolver, see~\autoref{sec:methodology}).
We then compare the per-protocol medians corresponding to a pair of \texttt{[vantage point:resolver]} to each other.
The relative differences to DoUDP (baseline) are shown in~\autoref{fig:udp-baseline}.

\begin{figure*}[t]
	\centering
	\includegraphics[width=\textwidth]{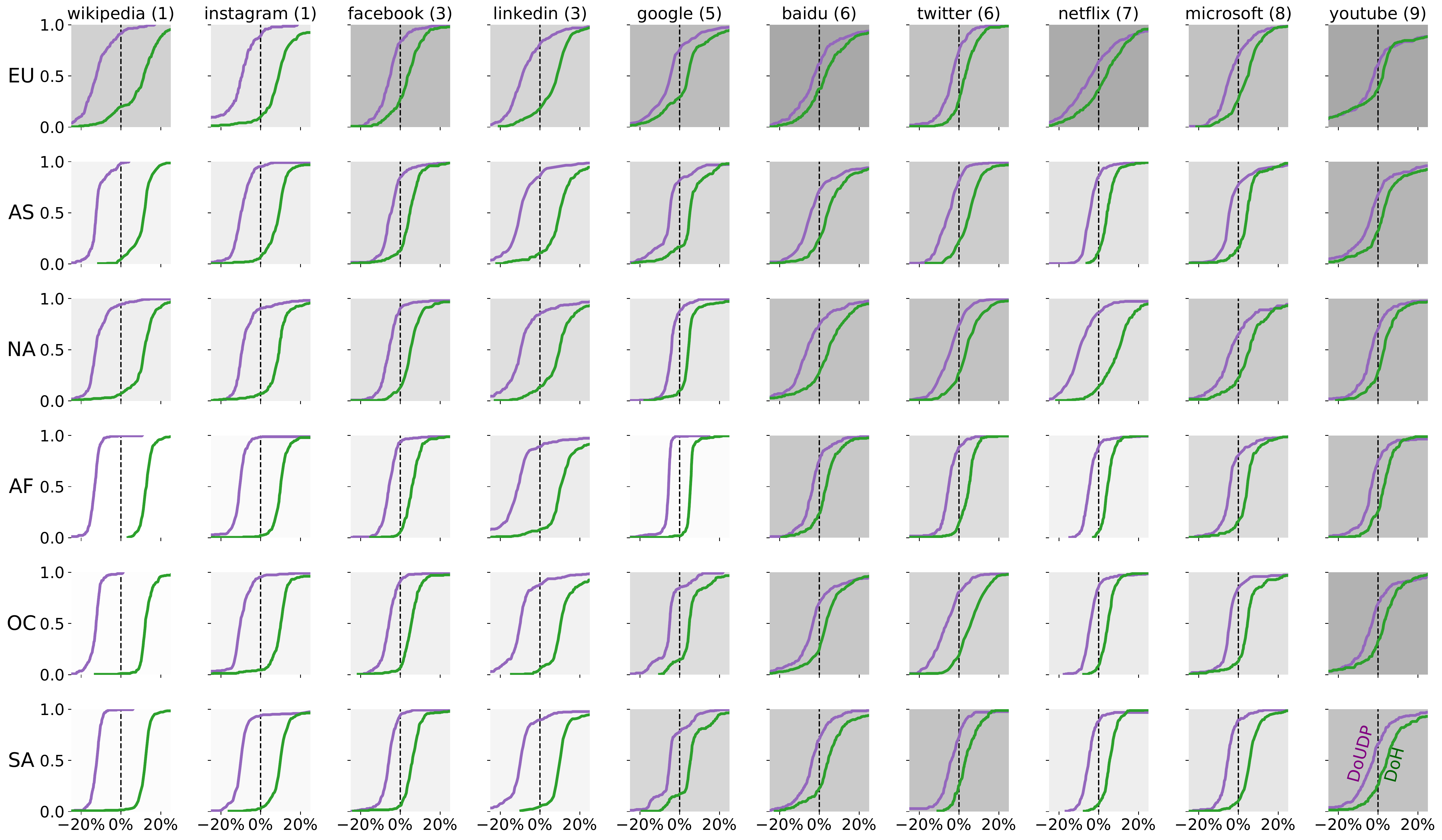}
	\vspace{-2em}
	\captionof{figure}{CDFs of the relative differences in PLT between DoQ (horizontal baseline), DoUDP (purple line), and DoH (green line), grouped by vantage point and webpage. A lighter background color depicts a higher percentage of DoQ page loads being faster than DoH. Sorted from left to right by the average number of DNS queries required for loading each webpage (in brackets) and from top to bottom by the number of verified DoX resolvers per continent.}
	\label{fig:doh-baseline-cluster}
\end{figure*}




\afblock{FCP.}
In almost 40\% of cases, using DoQ (\autoref{fig:udp-baseline-fcp}, blue line) delays the \acs{FCP} by 10\% or less when compared to DoUDP.
On the other hand, DoT (orange line) and DoH (green line) delay it by more than 20\% for the same fraction.
Looking at the 80th percentile, in 20\% of cases the \acs{FCP} increases by more than 20\% with DoQ, and almost twice as much with DoH.
In summary, DoQ performs much better than DoH and DoT when comparing the \acs{FCP} timings.

While both DoQ and DoTCP utilize a 1 round-trip connection setup (see~\autoref{sec:response-times}), we find that DoTCP is considerably slower than DoQ (\autoref{fig:udp-baseline-fcp}, red and blue lines).
Given that no resolver supports \texttt{edns-tcp-keepalive} or \acs{TFO} (see~\autoref{sec:evaluation}), DoTCP initiates a new connection for every query.
Hence, no resolver follows the recommendations for DoTCP~\cite{rfc9210} and each query takes 2 round-trips regardless of a previous connection establishment with the resolver.

In almost 10\% of cases, we find that the \acs{FCP} occurs faster when using DNS protocols other than DoUDP.
Analyzing this observation, we find that DoUDP is skewed by outliers: While both TCP and QUIC offer transport layer retransmission with initial timeouts of 1 second~\cite{rfc6298,rfc9002}, DoUDP is dependent on \emph{Chromium} retransmitting the DNS query on the application layer which uses a default initial timeout of 5 seconds~\cite{resolv.manpage}.

\afblock{PLT.}
The PLT presented in \autoref{fig:udp-baseline-plt} confirms our expectations that the relative impact of the DNS protocol is lower for \acs{PLT} than for the \acs{FCP}.
Overall, \ac{DoQ} (blue line) shows the smallest degradation in comparison to DoUDP, where less than 15\% of page loads increase the \acs{PLT} by more than 15\%.
In contrast, for more than 40\% of \ac{DoH} page loads, the \acs{PLT} also increases by more than 15\% (green line).
We find that the vast majority of the \ac{DoH} worst cases result from \emph{\nolinkurl{wikipedia.org}}, \emph{\nolinkurl{linkedin.com}}, and \emph{\nolinkurl{instagram.com}}; since those pages load very fast (due to the landing page mainly consisting of a login or search form), the impact of the DNS protocol on \acs{PLT} can still be relatively large.


Analyzing DoT (orange line), we find that it performs worse than DoH (green line).
While both protocols are expected to re-use previously established connections for subsequent DNS queries independent of the support of \texttt{edns-tcp-keepalive}~\cite{rfc7858,rfc8484}, a root cause analysis revealed that the connection handling in \emph{DNS Proxy} results in DoT repeating the full transport and encryption handshake in almost 60\% of page loads: While a DoT query is currently in-flight and a new request is issued, \emph{DNS Proxy} opens a new connection instead of re-using the existing one.
Hence, we disregard DoT in the following discussion and address this issue as part of our community contributions (see~\autoref{sec:methodology}).

\afblock{DoQ vs. DoH.}
We now take a closer look at DoQ and DoH to evaluate the performance differences of the encrypted DNS protocols in more detail.
For this, we analyze the impact of the vantage point and webpage on \acs{PLT}, where we present the relative differences between DoQ (horizontal baseline) and DoH (green line) in \autoref{fig:doh-baseline-cluster}.
Webpages are sorted from left to right by the average number of DNS queries required for loading each webpage (see columns); a lighter background color depicts a higher percentage of DoQ page loads being faster than DoH.

Overall, we find that DoQ mostly improves on DoH in all vantage point and webpage combinations, where the performance improvement diminishes the more DNS queries are required for loading a webpage (from left to right).
Analyzing the more simple webpages \emph{\nolinkurl{wikipedia.org}} and \emph{\nolinkurl{instagram.com}}, we observe that DoQ improves the PLT over DoH by up to 10\% in the median.
Hence, the simple webpages profit the most from \ac{DoQ}'s 1 round-trip connection establishment (see~\autoref{sec:response-times}), as there is only one DNS query on average.

Analyzing the differences between vantage points, we find that \ac{EU} shows the smallest differences between \ac{DoQ} and \ac{DoH} (top row).
Since we also observed the lowest response times on that vantage point (see~\autoref{sec:response-times}), we suspect that lower response times result in a smaller influence of the DNS protocol on the \acs{PLT}.
We find that the other vantage points also align in that trend, where DoQ positively impacts the performance more often when the response times are larger.
Nonetheless, we can already see an effect even with moderate response times, \eg in around 50\% of measurements, \emph{\nolinkurl{linkedin.com}} is more than 10\% faster with DoQ than with DoH in \ac{AS}.
However, we cannot determine a linear correlation between DNS protocol and \acs{PLT}.

\afblock{DoQ vs. DoUDP.}
Because unencrypted DNS traffic is still the norm, we next analyze the performance difference between DoQ and DoUDP (\autoref{fig:doh-baseline-cluster}, horizontal baseline and purple line).
Expectedly, DoUDP shows a better performance than DoQ in almost every case.
However, we also find DoUDP being slower than DoQ in some cases in the long tail, which is more pronounced in \ac{EU} (top row) as well as for webpages with a higher number of DNS queries.
As discussed for \emph{FCP}, we attribute this observation to DoUDP being skewed by outliers due to application layer retransmissions.

For the more simple webpages \emph{\nolinkurl{wikipedia.org}} and \emph{\nolinkurl{instagram.com}}, we find that the performance cost of encryption is the largest with up to 10\% slower PLT in the median due to the added overhead of \ac{DoQ}'s connection establishment.
However, we find that the difference in PLT is reduced to only $\sim$2\% in the median between DoQ and DoUDP for the more complex webpages \emph{\nolinkurl{microsoft.com}} and \emph{\nolinkurl{youtube.com}}: DoQ even catches up to DoUDP as the encryption overhead amortizes the more DNS queries are required for loading a webpage.

\takeaway{
DoQ significantly improves over DoH.
While we find that page loads using DoQ are up to 10\% faster for simple webpages in comparison to DoH, the cost of encryption is the largest for the same webpages, where DoQ is up to 10\% slower than DoUDP.
With increasing complexity of webpages, however, DoQ catches up to DoUDP as the cost of encryption amortizes the more DNS queries are required for loading a webpage: DoQ is only $\sim$2\% slower than DoUDP, thus, making encrypted DNS much more appealing for the Web.
}

\section{Limitations and Future Work}
\label{sec:limitations}

\afblock{Limitations.}
Note that the Web performance measurements only consider a total number of 10 webpages, which might not be representative for the Web as a whole; with an increased number of requests per page and increased webpage complexity, the benefits of DoQ in comparison to DoH (e.g., fewer round-trips required for the handshake) are diminished due to amortization and other confounding factors such as webpage rendering.
Further, considering the limited number of 313 DoX resolvers which are heavily centered around Europe, some vantage points experience higher latency due to larger geographical distances to the targeted resolvers.
While we briefly discuss resulting outliers in the previous sections, a detailed root cause analysis, esp. for the Web performance measurements, is left for future work.


\afblock{Future Work.}
With the ongoing development and adoption of DoQ among resolvers, we expect resolvers to introduce support for 0-RTT in the future, which can shift the total response times of DoQ even closer to DoUDP.
Thus, we plan to continue measuring and monitoring the rollout of DoQ.
Moreover, while the DoH measurements in our study use HTTP/2 (see~\autoref{sec:evaluation}), we will extend our work with an in-depth comparison to DNS over HTTP/3 (DoH3): The recently standardized HTTP/3~\cite{rfc9114} also uses QUIC as its transport protocol.
At the time of writing, DoH3 is not yet widely supported:
While Cloudflare is one of the first to support DoH3~\cite{cloudflare-ddr} by including HTTP/3 in the ALPN set of their \texttt{SVCB} records~\cite{ietf-dnsop-svcb-https}, we observe that state of the art browsers only connect to Cloudflare's resolvers via HTTP/2, which indicates that DoH3 support among browsers is still lacking.
However, with its recent integration into Google Public DNS and Android~\cite{android-doh3}, DoH3 is expected to gain momentum in the coming months.


\section{Conclusion}
\label{sec:conclusion}

Our study showed that encrypted DNS does not have to be a compromise between privacy and speed: Using DoQ, the single query response time is improved by $\sim$33\% in comparison to DoT and DoH.
The Web performance measurements revealed that DoQ significantly improves over DoH with up to 10\% faster loads for simple webpages.
With increasing complexity of webpages, DoQ even catches up to DoUDP as the cost of encryption amortizes: With DoQ being only $\sim$2\% slower than DoUDP, encrypted DNS becomes much more appealing for the Web, especially once resolvers start supporting advanced features such as 0-RTT.

\section*{Acknowledgements}
We thank Justus Fries, Sebastian Kappes, and Malte Granderath for their valuable support, as well as Jan Rüth, the anonymous reviewers, and our shepherd Stephen McQuistin for their insightful feedback.
This work was partly supported by the Volkswagenstiftung Niedersächsisches Vorab (Funding No. ZN3695).

\bibliographystyle{ACM-Reference-Format}
\bibliography{references}

\raggedend

\end{document}